\documentclass[12pt]{article}
\usepackage{graphicx,xcolor}
\usepackage{axodraw,epsf}

\newcommand{\beq}{\begin{equation}}
\newcommand{\eeq}{\end{equation}}
\newcommand{\bea}{\begin{eqnarray}}
\newcommand{\eea}{\end{eqnarray}}

\begin{document}
\begin{titlepage}
\begin{center}
{\LARGE \bf  The $\gamma^* \gamma^*$ total cross section in
next-to-leading

order BFKL and LEP2 data}
\end{center}

\vskip 0.5cm

\centerline{D.Yu.~Ivanov$^{1\P}$, B.~Murdaca$^{2\dag}$ and
A.~Papa$^{2\ddagger}$}

\vskip .6cm

\centerline{${}^1$ {\sl Sobolev Institute of Mathematics and Novosibirsk State University,}}
\centerline{\sl 630090 Novosibirsk, Russia}

\vskip .2cm

\centerline{${}^2$ {\sl Dipartimento di Fisica, Universit\`a della Calabria,}}
\centerline{\sl and Istituto Nazionale di Fisica Nucleare, Gruppo collegato di
Cosenza,}
\centerline{\sl Arcavacata di Rende, I-87036 Cosenza, Italy}

\vskip 2cm

\begin{abstract}
We study the total cross section for the collision of two highly-virtual
photons at large energies, taking into account the BFKL resummation of energy
logarithms with full next-to-leading accuracy. A necessary ingredient of the
calculation, the next-to-leading order impact factor for the photon to
photon transition, has been calculated by Balitsky and Chirilli using an
approach based on the operator expansion in Wilson lines.
We extracted the result for the photon impact factor in the original BFKL
calculation scheme comparing the expression for the photon-photon total
cross section obtained in BFKL with the one recently derived
by Chirilli and Kovchegov in the Wilson-line operator expansion scheme.

We perform a detailed numerical analysis, combining different, but
equivalent in next-to-leading accuracy, representations of the cross section
with various optimization methods of the perturbative series. We compare our
results with previous determinations in the literature and with the
LEP2 experimental data. We find that the account of Balitsky and Chirilli
expression  for the photon impact factor reduces the BFKL contribution to the
cross section to very small values, making it impossible to describe LEP2
data as the sum of BFKL and leading-order QED quark box contributions.
\end{abstract}


$
\begin{array}{ll} ^{\P}\mbox{{\it e-mail address:}} &
\mbox{d-ivanov@math.nsc.ru}\\
^{\dag}\mbox{{\it e-mail address:}} &
\mbox{beatrice.murdaca@fis.unical.it}\\
^{\ddagger}\mbox{{\it e-mail address:}} &
\mbox{alessandro.papa@fis.unical.it}\\
\end{array}
$

\end{titlepage}

\vfill \eject

\section{Introduction}

\begin{figure}[tb]
\begin{minipage}{70mm}
\begin{center}
\setlength{\unitlength}{0.35mm}
\begin{picture}(150,150)(0,0)
\Photon(10,140)(50,100){3}{7}
\ArrowLine(50,100)(100,100)
\Photon(100,100)(140,140){3}{7}
\ArrowLine(50,100)(50,50)
\Photon(10,10)(50,50){3}{7}
\ArrowLine(50,50)(100,50)
\Photon(100,50)(140,10){3}{7}
\ArrowLine(100,50)(100,100)
\end{picture}
\end{center}
\end{minipage}
\begin{minipage}{90mm}
\begin{center}
\setlength{\unitlength}{0.35mm}
\begin{picture}(150,150)(0,0)
\Photon(10,140)(50,100){3}{7}
\Line(50,100)(100,50)
\Photon(100,100)(140,140){3}{7}
\ArrowLine(50,100)(50,50)
\Photon(10,10)(50,50){3}{7}
\Line(50,50)(100,100)
\Photon(100,50)(140,10){3}{7}
\ArrowLine(100,50)(100,100)
\end{picture}
\end{center}
\end{minipage}
\caption[]{Quark box LO diagrams.}
\label{qbox}
\end{figure}
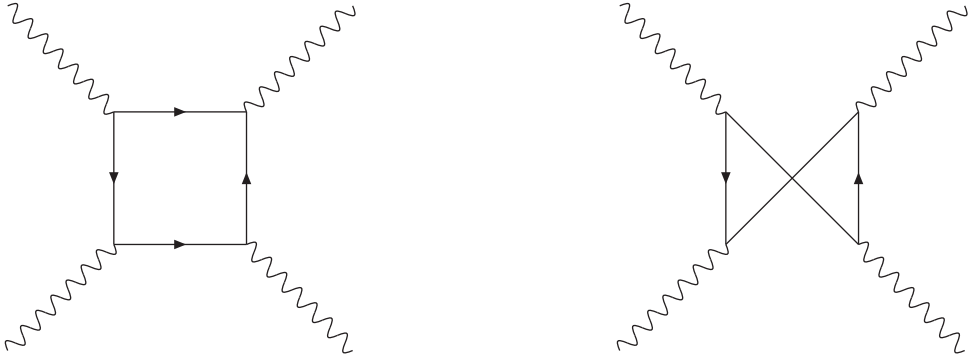

Similarly to the $e^+e^-$ annihilation into hadrons, the total cross section
for the collision of two off-shell photons with large virtualities is an
important test ground for perturbative QCD.
At a fixed order of $\alpha_s$ and at low energies, the dominant contribution
comes from the quark box, calculated at the leading-order (LO)
in Refs.~\cite{Budnev:1974de,Schienbein:2002wj} (see Fig.~\ref{qbox}) and
at the next-to-LO (NLO) in Ref.~\cite{Cacciari:2000cb}.
In Ref.~\cite{BL03} the resummation of double logs appearing in the NLO
corrections to the quark box was also studied.
At higher energies, the gluon exchange in the $t$-channel overwhelms the
quark exchange contribution, due to the different power asymptotics
for $s\to \infty$. At higher orders in $\alpha_s$, the contributions
from $t$-channel gluons lead to terms with powers of single logarithms of
the energy, which must be resummed.

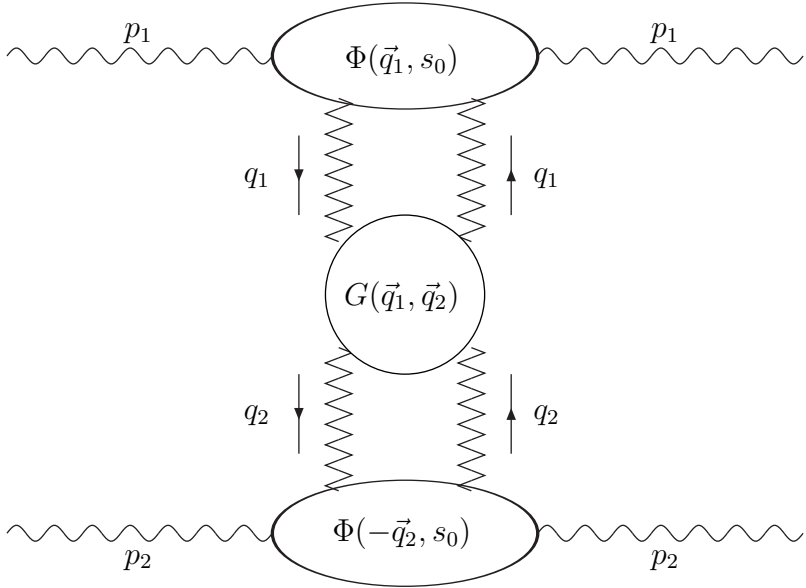
\begin{figure}[tb]
\centering
\setlength{\unitlength}{0.35mm}
\begin{picture}(300,200)(0,0)

\Photon(0,190)(100,190){3}{7}
\Photon(200,190)(300,190){3}{7}
\Text(50,200)[c]{$p_1$}
\Text(250,200)[c]{$p_1$}
\Text(150,190)[]{$\Phi(\vec q_1, s_0)$}
\Oval(150,190)(20,50)(0)

\ZigZag(125,174)(125,120){5}{7}
\ZigZag(175,174)(175,120){5}{7}
\ZigZag(125,26)(125,80){5}{7}
\ZigZag(175,26)(175,80){5}{7}

\ArrowLine(110,160)(110,130)
\ArrowLine(190,130)(190,160)
\ArrowLine(110,70)(110,40)
\ArrowLine(190,40)(190,70)

\Text(100,145)[r]{$q_1$}
\Text(200,145)[l]{$q_1$}
\Text(100,55)[r]{$q_2$}
\Text(200,55)[l]{$q_2$}

\GCirc(150,100){30}{1}
\Text(150,100)[]{$G(\vec q_1,\vec q_2)$}

\Photon(0,10)(100,10){3}{7}
\Photon(200,10)(300,10){3}{7}
\Text(50,0)[c]{$p_2$}
\Text(250,0)[c]{$p_2$}
\Text(150,10)[]{$\Phi(-\vec q_2,s_0)$}
\Oval(150,10)(20,50)(0)

\end{picture}

\caption[]{Schematic representation of the elastic amplitude for the
$\gamma^*(p_1)\, \gamma^*(p_2)$ forward scattering.}
\label{fig:BFKL}
\end{figure}

The BFKL approach~\cite{BFKL} provides for a consistent theoretical framework
for the resummation of the energy logarithms, both in the leading logarithmic
approximation (LLA), which means resummation of all terms
$(\alpha_s\ln(s))^n$, and in the next-to-leading approximation (NLA), which
means resummation of all terms $\alpha_s(\alpha_s\ln(s))^n$. In this
approach, the imaginary part of the amplitude (and, hence, the total cross
section) for a large-$s$ hard collision process can be written as
the convolution of the Green's function of two interacting Reggeized gluons
with the impact factors of the colliding particles (see Fig.~\ref{fig:BFKL}).

The study of the $\gamma^* \gamma^*$ total cross section in LLA BFKL has a long 
history~\cite{photons_BFKL}. For the extension of these results to the NLA 
level one needs to consider corrections to both the BFKL Green's function and 
to the impact factors of colliding virtual photons.

The Green's function is determined through the BFKL equation and is
process-inde\-pen\-dent. The NLO kernel of the BFKL equation for singlet color
representation in the $t$-channel and forward scattering, relevant for
the determination of a total cross section in the NLA, has been achieved in
Refs.~\cite{NLA-kernel}, after the long program of calculation
of the NLO corrections~\cite{NLA-corrections} (for a review, see
Ref.~\cite{news}).

The other essential ingredient for the $\gamma^* \gamma^*$ total cross
section is the impact factor for the virtual photon to virtual photon
transition. While its LO expression is known since long, the NLO
calculation, carried out in the momentum representation, turned out to be
rather complicated and was completed only after year-long
efforts~\cite{gammaIF}. The lengthy result was published over a few years in
pieces, some of them available only in the form of a numerical code,
thus making it of limited practical use. Indeed, until very recently,
the inclusion of BFKL resummation effects in the NLA calculation of the
$\gamma^* \gamma^*$ total cross section was carried out only in approximate
way, by taking the BFKL Green's function in the NLA while using the LO
expression for impact factors. This is the case of the pioneer paper in
Ref.~\cite{Brodsky:2002ka} (see also Ref.~\cite{Brodsky:1998kn}) and of the
later analysis in Refs.~\cite{Caporale2008} and~\cite{Zheng}.

The situation changed radically recently, when the NLO photon impact
was calculated in the coordinate space and then transformed to
the momentum representation and to the Mellin (or
$\gamma$-representation)~\cite{Balitsky2012} (see also
Ref.~\cite{Chirilli2014}). The NLO expression for the photon impact factor
turns out to be very simple in all representations, thus confirming an
already well established evidence (see, for instance,
Refs.~\cite{coordinate}) that the use of the coordinate representation leads
to much simpler expressions for the NLO BFKL kernel and impact factors,
which, in the momentum representation, would be the result of not so obvious
cancellations.

Now all ingredients are available to build the $\gamma^* \gamma^*$ total
cross with {\em full} inclusion of the BFKL resummation in the NLA. Indeed,
already in Ref.~\cite{Chirilli2014} there is a first numerical estimate of
the $\gamma^* \gamma^*$ total cross section in the NLA. Note that, the
derivation of the results for the $\gamma^* \gamma^*$ total cross section in
Ref.~\cite{Chirilli2014} follows closely the approach developed earlier in
Ref.~\cite{Balitsky:2009yp}, where the high-energy limit of ${\cal N}=4$ SYM
amplitudes was considered. Besides, the authors of Ref.~\cite{Chirilli2014}
used their formulas for the eigenfunctions of the NLO BFKL kernel
derived in Ref.~\cite{Chirilli:2013kca}.

As a matter of fact, previous studies of physical processes within the BFKL
approach in the NLA, such as the photoproduction of two light vector
mesons~\cite{IKP04,mesons_1-2,mesons_3} and the production of Mueller-Navelet
jets~\cite{MN_IF,MN_sigma}, have clearly shown that NLA expressions for
an observable (such as a cross-section or an azimuthal correlation), though
being formally equivalent up to subleading terms, may lead to somewhat
different numerical estimates.
At the basis of this observation is the fact that NLO BFKL corrections, both
of the kernel and of impact factors, are typically of opposite sign with
respect to the LO and large in absolute value. For this reason a numerical
estimate cannot be reliable (i) if some optimization procedure for the
perturbative series is not applied and (ii) if not corroborated by a careful
numerical analysis, aimed at assessing the stability of the result under
variation of the original NLA expressions for the observable of interest
within a large enough class of NLA-equivalent expressions.

The aim of this paper is to contribute to such analysis, by comparing several
NLA-equivalent representations of the $\gamma^* \gamma^*$ total cross section,
in combination with two among the most common methods of optimization of the
perturbative series, namely the principle of minimal sensitivity
(PMS)~\cite{Stevenson} and the Brodsky-Lepage-Mackenzie (BLM)
method~\cite{BLM}. Moreover, the results of this analysis will be contrasted
with the only experimental data available so far, obtained at
LEP2~\cite{Achard:2001kr,Abbiendi:2001tv}.

The paper is organized as follows: In Section~2 we present the general
structure of the $\gamma^* \gamma^*$ total cross section in the NLA and,
by comparison with Refs.~\cite{Balitsky2012,Chirilli2014}, extract the NLO
photon impact factor in the original BFKL calculation scheme; in Section~3 we
use this information to build
several NLA-equivalent representations of the cross section and present,
for each of them, the behavior with the energy in comparison with the LEP2
experimental data; finally, in Section~4, we discuss our results and draw
the conclusions.

\section{BFKL contribution to the $\gamma^* \gamma^*$ total cross section}

The total cross section of two unpolarized photons with virtualities
$Q_1$ and $Q_2$ can be obtained from the imaginary part of the forward
amplitude. In LLA BFKL and in the Mellin-representation (also
said $\gamma$- or $\nu$-representation), it is given by the following
expression (see, for instance, Ref.~\cite{Brodsky:2002ka}):
\begin{equation}
\sigma^{\gamma^{*} \gamma^{*}}_{\rm tot} (s,Q_1,Q_2) =
\sum_{i,k=T,L} \frac{1}{(2 \pi)^2 Q_1 Q_2}
\int\limits^{+\infty}_{-\infty} d\nu \left(\frac{Q_1^2}{Q_2^2}\right)^{i\nu}
F_i(\nu) F_k(-\nu) \left(\frac{s}{s_0}\right)^{\bar \alpha_s \chi(\nu)} \; ,
\label{sigmaLO}
\end{equation}
where $\bar \alpha_s\equiv \alpha_s(\mu_R) N_c/\pi$, with $N_c$ the number
of colors, $\chi(\nu)$ is the so-called characteristic BFKL function,
\beq
\chi(\nu)=2\psi(1)-\psi\left(\frac{1}{2}+i\nu\right)-\psi\left(\frac{1}{2}
-i\nu\right)
\label{chi}
\eeq
and
\bea
F_T(\nu) = F_T(-\nu)&=& \alpha \, \alpha_s \left( \sum_q e_q^2 \right)
\frac{\pi}{2} \frac{(\frac{3}{2} - i\nu)(\frac{3}{2}+ i\nu)
\Gamma^2(\frac{1}{2} - i\nu)\Gamma^2(\frac{1}{2}+i\nu)}
{\Gamma(2-i\nu) \Gamma(2+i\nu)}
\nonumber \\
&=& \alpha \, \alpha_s \left( \sum_q e_q^2 \right)
\frac{\pi^2}{8}\frac{9+4\nu^2}{\nu\left(1+\nu^2\right)}
\frac{\sinh \left(\pi\nu\right)}{\cosh^2(\pi\nu)} \, , \label{FT_LO}
\eea
\bea
F_L(\nu) = F_L(-\nu)&=& \alpha \, \alpha_s \left( \sum_q e_q^2 \right) \pi
\frac{\Gamma(\frac{3}{2} - i\nu)\Gamma(\frac{3}{2} + i\nu)
\Gamma(\frac{1}{2} - i\nu) \Gamma(\frac{1}{2} + i\nu)}
{\Gamma(2-i\nu) \Gamma(2+i\nu)} \nonumber \\
&=& \alpha \, \alpha_s \left( \sum_q e_q^2 \right)
\frac{\pi^2}{4}\frac{1+4\nu^2}{\nu\left(1+\nu^2\right)}
\frac{\sinh \left(\pi\nu\right)}{\cosh^2(\pi\nu)} \label{FL_LO}
\eea
are the LO impact factors for transverse and longitudinal polarizations,
respectively. In the previous equations, $\alpha$ is the electromagnetic
coupling constant, the summation extends over all active quarks (taken
massless) and $e_q$ is the quark electric charge in units of the electron
charge. In the expression~(\ref{sigmaLO}) for the LLA BFKL cross section the
argument of the strong and electromagnetic coupling constants, $\mu_R$,
and the value of the scale $s_0$ are not fixed.

Following the procedure of Refs.~\cite{mesons_1-2}, it is possible
to write down the NLA BFKL cross section as follows:
\[
\sigma^{\gamma^{*} \gamma^{*}}_{\rm tot} (s,Q_1,Q_2,s_0,\mu_R)
= \frac{1}{(2 \pi)^2 Q_1 Q_2}
\int\limits^{+\infty}_{-\infty} d\nu \left(\frac{Q_1^2}{Q_2^2}\right)^{i\nu}
\left(\frac{s}{s_0}\right)^{\bar \alpha_s(\mu_R) \chi(\nu)}
\]
\beq
\times\sum_{i,k=T,L}F_i(\nu)F_k(-\nu) \left\{1+\bar\alpha_s(\mu_R)
\left(\frac{F_i^{(1)}(\nu,s_0,\mu_R)}{F_i(\nu)}+\frac{F_k^{(1)}(-\nu,s_0,\mu_R)}{F_k(-\nu)}\right)
\right. \label{sigmaNLO}
\eeq
\[
\left.
+ \bar \alpha_s^2(\mu_R) \ln\left(\frac{s}{s_0}\right) \left[ \bar
\chi(\nu)+\frac{\beta_0}{8N_c}\chi(\nu)\left(-\chi(\nu)+\frac{10}{3}
+ 2\ln\frac{\mu_R^2}{Q_1Q_2} \right) \right]\right\} \; ,
\]
where
\bea
\bar\chi(\nu)\,&=&\,-\frac{1}{4}\left[\frac{\pi^2-4}{3}\chi(\nu)-6\zeta(3)-
\chi^{\prime\prime}(\nu)-\frac{\pi^3}{\cosh(\pi\nu)}
\right.
\nonumber \\
&+& \left.
\frac{\pi^2\sinh(\pi\nu)}{2\,\nu\, \cosh^2(\pi\nu)}
\left(
3+\left(1+\frac{n_f}{N_c^3}\right)\frac{11+12\nu^2}{16(1+\nu^2)}
\right)
+\,4\,\phi(\nu)
\right] \, ,
\eea
\beq
\phi(\nu)\,=\,2\int\limits_0^1dx\,\frac{\cos(\nu\ln(x))}{(1+x)\sqrt{x}}
\left[\frac{\pi^2}{6}-\mbox{Li}_2(x)\right]\, , \;\;\;\;\;
\mbox{Li}_2(x)=-\int\limits_0^xdt\,\frac{\ln(1-t)}{t} \, ,
\eeq
$n_f$ is the number of active quarks, $F^{(1)}_{L,T}(\nu,s_0,\mu_R)$ are the
NLO corrections to the longitudinal/transverse photon impact factor in the
$\nu$-representation and
\beq
\beta_0= \frac{11}{3}N_c-\frac{2}{3}n_f\; .
\eeq

Note that our notations are slightly different in comparison to the ones
used in Refs.~\cite{mesons_1-2}. The impact factors which we introduced here
differ by some factors from the impact factors $c_{1,2}$ (and $c^{(1)}_{1,2}$)
of Refs.~\cite{mesons_1-2}:
\bea
& c_{1,i}(\nu)=\left(Q^2_1\right)^{i\nu-1/2}F_i(\nu) & \nonumber \\
& c_{2,k}(\nu)=\left(Q^2_2\right)^{-i\nu-1/2}F_k(-\nu) & \nonumber \, .
\eea
Moreover, in the derivation of the last term of Eq.~(\ref{sigmaNLO}) the
symmetry property of the LO photon impact factors,
$F_{L,T}(\nu)=F_{L,T}(-\nu)$, was used.

Our goal now is to extract the NLO parts of the photon impact factors,
$F^{(1)}_{L,T}(\nu,s_0,\mu_R)$, which enter the cross section
Eq.~(\ref{sigmaNLO}) in the original BFKL approach, by comparing
Eq.~(\ref{sigmaNLO}) with the results for the $\gamma^*\gamma^*$ cross section
obtained recently in  the Wilson-line operator expansion scheme by Chirilli
and Kovchegov~\cite{Chirilli2014}.
According to Eqs.~(3.40) and~(3.41) of Ref.~\cite{Chirilli2014}, the cross
section in the case of transverse and longitudinal polarizations reads
\begin{eqnarray}
\label{sigmaNLO-T-CK}
\sigma^{\rm (CK)}_{TT}
&=&\left( \sum_q e_q^2\right)^2\frac{\alpha^2
\alpha_s^2}{ Q_1 Q_2}\frac{\pi^2}{2^8} \int_{-\infty}^{+\infty}
d\nu \left(\frac{Q_1^2}{Q_2^2}\right)^{i\nu}
\left(\frac{s}{Q_1 Q_2}\right)^{\bar\alpha_s
\chi\left(\nu\right)+\bar\alpha_s^2\chi^{\left(1\right)}
\left(\nu\right)}
\nonumber \\
&\times&\left[\frac{\left(9+4\nu^2\right)}{\nu\left(1+\nu^2\right)}
\frac{\sinh\left(\pi\nu\right)}{\cosh^2(\pi\nu)}\right]^2
\left[1+\frac{\alpha_s}{\pi}+\frac{\bar\alpha_s}{2}{\cal F}_1
\left(\nu\right)\right]\left[1+\frac{\alpha_s}{\pi}+\frac{\bar\alpha_s}{2}
{\cal F}_1\left(-\nu\right)\right] \nonumber \\
&\times&\left\lbrace1+\bar\alpha_s\Re\left[F\left(\nu\right)\right]
\right\rbrace\;,
\end{eqnarray}

\begin{eqnarray}
\label{sigmaNLO-L-CK}
\sigma^{\rm (CK)}_{LL}
&=&\left(\sum_q e_q^2\right)^2\frac{\alpha^2
\alpha_s^2}{Q_1 Q_2}\frac{\pi^2}{2^8}
\int_{-\infty}^{+\infty} d\nu \left(\frac{Q_1^2}{Q_2^2}\right)^{i\nu} \left(\frac{s}{Q_1 Q_2}\right)^{\bar\alpha_s
\chi\left(\nu\right)+\bar\alpha_s^2
\chi^{\left(1\right)}\left(\nu\right)} \nonumber \\
&\times&\left[\frac{\left(9+4\nu^2\right)}{\nu\left(1+\nu^2\right)}\frac{\sinh\left(\pi\nu\right)}
{\cosh^2(\pi\nu)}\right]^2 \left[\frac{11+12 \nu^2}{9+4 \nu^2}\left(1+\frac{\alpha_s}
{\pi}+\frac{\bar\alpha_s}{2}{\cal F}_2(\nu)\right)\right. \nonumber \\
&-&\left. \left( 1+\frac{\alpha_s}{\pi}+\frac{\bar\alpha_s}{2}{\cal F}_1\left(\nu\right)\right)
\right]\left[\frac{11+12 \nu^2}{9+4 \nu^2}\left(1+\frac{\alpha_s}
{\pi}+\frac{\bar\alpha_s}{2}{\cal F}_2(-\nu)\right)\right.
\nonumber \\
&-&\left. \left( 1+\frac{\alpha_s}{\pi}+\frac{\bar\alpha_s}{2}{\cal F}_1\left(-\nu\right)\right)
\right]\left\lbrace1+\bar\alpha_s\Re\left[F\left(\nu\right)\right]
\right\rbrace\;,
\end{eqnarray}
where in the r.h.s. the strong coupling without argument stands
for the coupling at the symmetric point, $\alpha_s=\alpha_s(\sqrt{Q_1 Q_2})$,
$\Re\left[F\left(\nu\right)\right]$ is given in Eq.~(3.37) of
Ref.~\cite{Chirilli2014} and the explicit expressions for ${\cal F}_{1,2}
\left(\nu\right)$ in Eq.~(52) of  Ref.~\cite{Balitsky2012};
for the definition of $\chi^{\left(1\right)}\left(\nu\right)$, see Eqs.~(2.9)
and~(2.11) of Ref.~\cite{Chirilli2014}, so that
\beq
\chi^{\left(1\right)}\left(\nu\right)=\bar
\chi(\nu)+\frac{\beta_0}{8N_c}\chi(\nu)\left(-\chi(\nu)+\frac{10}{3}
 \right) \; .
\eeq

Note that working with NLA accuracy, we replaced in Eqs.~(\ref{sigmaNLO-T-CK})
and~(\ref{sigmaNLO-L-CK}) the original factors
$\alpha_s\left(Q_1\right)\alpha_s\left(Q_2\right)$ present in Eqs.~(3.40)
and~(3.41) of Ref.~\cite{Chirilli2014} by $\alpha_s^2(\sqrt{Q_1 Q_2})$, since
$$\alpha_s\left(Q_1\right)\alpha_s\left(Q_2\right)
=\alpha_s^2(\sqrt{Q_1 Q_2})+{\cal O}(\alpha_s^4)\;.$$
Another point is that the terms $\alpha_s/\pi$ which appear in
the r.h.s. of Eqs.~(\ref{sigmaNLO-T-CK}) and~(\ref{sigmaNLO-L-CK}) are due to
the QCD vacuum polarization contribution, which actually reads as
$3 C_F\alpha_s/(4\pi)$, where
$C_F=(N_c^2-1)/(2 N_c)$ \footnote{We are very grateful to the authors of
Ref.~\cite{Chirilli2014} for the clarification of this issue and for
establishing the overall normalization factor in their results for the
cross section.}.

Now we are ready to compare Eqs.~(\ref{sigmaNLO-T-CK})
and~(\ref{sigmaNLO-L-CK}) with the BFKL cross section Eq.~(\ref{sigmaNLO})
taken for the particular choice of scales $\mu_R^2=s_0=Q_1 Q_2$.
Expanding Eqs.~(\ref{sigmaNLO-T-CK}) and~(\ref{sigmaNLO-L-CK}) into a form
similar to Eq.~(\ref{sigmaNLO}) and requiring the coincidence of the
two representations for the cross section with NLA accuracy allows us to
extract without ambiguity the NLO parts of the BFKL impact factors
$F^{(1)}_{L,T}(\nu,s_0,\mu_R)$ (at the scale setting $s_0=\mu_R^2=Q_1 Q_2$):

\bea
\label{ifT}
\frac{F_T^{(1)}(\nu,s_0,\mu_R)}{F_T(\nu)}
&=&
\frac{\chi(\nu)}{2}\ln\frac{s_0}{Q^2}+\frac{\beta_0}{4 N_c}\ln\frac{\mu_R^2}{Q^2}
 \\
&+&\frac{3 C_F}{4 N_c}-\frac{5}{18}\frac{n_f}{N_c}+\frac{\pi^2}{4}+\frac{85}{36}
-\frac{\pi^2}{\cosh^2(\pi\nu)}-\frac{4}{1+4\nu^2}
+\frac{6\chi\left(\nu\right)}{9+4\nu^2}
\nonumber \\
&+&\frac{1}{2\left(1-2 i\nu\right)}-\frac{1}{2\left( 1+2 i\nu\right)}
-\frac{7}{18\left( 3+2 i\nu\right)}+\frac{20}{3\left(3+2 i\nu\right)^2}
-\frac{25}{18\left(3-2 i\nu\right)}
\nonumber
\\
&+&\frac{1}{2}\chi\left(\nu\right)\left[\psi\left(\frac{1}{2}-i\nu\right)
+2\psi\left(\frac{3}{2}-i\nu\right)-2\psi\left(3-2 i\nu\right)
-\psi\left(\frac{5}{2}+i\nu\right)\right] \nonumber
\eea
and

\bea
\label{ifL}
\frac{F_L^{(1)}(\nu,s_0,\mu_R)}{F_L(\nu)}
&=&\frac{\chi(\nu)}{2}\ln\frac{s_0}{Q^2}+\frac{\beta_0}{4 N_c}\ln\frac{\mu_R^2}{Q^2}
\\
&+&\frac{3 C_F}{4 N_c}-\frac{5}{18}\frac{n_f}{N_c}+\frac{\pi^2}{4}+\frac{85}{36}
-\frac{\pi^2}{\cosh^2(\pi\nu)}
-\frac{8\left(1+4 i\nu \right)}{\left( 1+2 i\nu \right)^2\left(1-2 i\nu \right)\left(3 +2 i\nu \right)}
\nonumber \\
&+&\frac{4}{3-4 i\nu +4\nu^2}
\chi\left(\nu\right)
\nonumber \\
&+&\frac{1}{2}\chi\left(\nu\right)\left[\psi\left(\frac{1}{2}-i\nu\right)
+2\psi\left(\frac{3}{2}-i\nu\right)-2\psi\left(3-2 i\nu\right)
-\psi\left(\frac{5}{2}+i\nu\right)\right]\;.
\nonumber
\eea
The first lines of Eqs.~(\ref{ifT}) and~(\ref{ifL}) describe the dependencies
of the photon impact factors on the renormalization and energy scales,
which are restored by the requirement that the BFKL cross section,
Eq.~(\ref{sigmaNLO}), does not depend on $s_0$ and $\mu_R$ with NLA accuracy.

\section{Numerical analysis}

In this Section we are going to compare several different representations
of the NLA $\gamma^* \gamma^*$ total cross section, which differ one from
the other only by terms beyond the NLA. In a well-behaved perturbative
series, the change of representation should not be numerically relevant.
This is not the case in the BFKL framework, where it is well known that
NLO corrections to kernel and impact factors are opposite in sign with respect
to the LO contributions and large in absolute value.

It is very likely that also the (unknown) next-to-NLO corrections
maybe opposite in sign with respect the NLO ones and large in absolute value,
thus suggesting that fixing the BFKL energy scale $s_0$ and the renormalization
scale $\mu_R$ at the ``natural'' values dictated by the kinematics of the
process, {\it i.e.} $s_0=\mu_R^2=Q_1 Q_2$, may well be not the best choice.
For this reason, we will consider in the following two alternative
procedures to fix the energy scales.

The first one is inspired by the PMS optimization method~\cite{Stevenson}:
for each value of the center-of-mass energy $s$ and of the virtualities
of the colliding photons, we choose as optimal scales $s_0$ and $\mu_R$
those for which the given representation of the NLA cross section exhibits
the minimum sensitivity under variation of these scales.

The other optimization procedure we consider is inspired by the BLM
method~\cite{BLM}: again, for fixed $s$ and photon virtualities, we
perform a finite renormalization to a momentum (MOM) scheme and then choose
the renormalization scale $\mu_R$ in order to remove the $\beta_0$-dependent
part in the given representation of the NLA cross section, while keeping
the scale $s_0$ fixed at the natural value $Q_1 Q_2$. In fact, there is some
freedom in implementing the BLM optimization in this context and in the
following we consider two different variants, dubbed $(a)$ and $(b)$, and
give all necessary formulas, but relegate their derivation to a separate
paper~\cite{us_BLM}.

Below we will present predictions for the kinematic range relevant for the
OPAL and L3 experiments at LEP2, considering equal photon virtualities,
$Q_1=Q_2\equiv Q$, with $Q^2$=17 GeV$^2$, and the energy range
$Y=2\div 6$, where $Y\equiv\ln(s/Q^2)$.

\subsection{Chirilli-Kovchegov representation}

As a first case, we try to apply to the description of LEP2 data the
representation of the NLA $\gamma^* \gamma^*$ total cross section
given in Ref.~\cite{Chirilli2014}. It is given, with obvious meaning of the
notation, by
\beq
\sigma^{\rm (CK)}_{\rm tot} (s,Q)
= \sigma^{\rm (CK)}_{TT} + \sigma^{\rm (CK)}_{LL}
+ \sigma^{\rm (CK)}_{TL} + \sigma^{\rm (CK)}_{LT} + \sigma_{\rm LO \ box}\;,
\label{sigmaCK}
\eeq
where we have included the LO contribution from the quark box,
given in Refs.~\cite{Budnev:1974de,Schienbein:2002wj}, and contributions of
different polarization states of virtual photons.  The explicit
expressions for the first two terms in Eq.~(\ref{sigmaCK}) are given in
Eqs.~(\ref{sigmaNLO-T-CK}) and~(\ref{sigmaNLO-L-CK}) (for $Q_1=Q_2= Q$), the
contributions of other polarizations can be obviously presented as follows:

\begin{eqnarray}
\sigma^{\rm (CK)}_{TL}
&=&\left( \sum_q e_q^2\right)^2\frac{\alpha^2
\alpha_s^2}{Q^2}\frac{\pi^2}{2^8} \int_{-\infty}^{+\infty} d\nu
\left(\frac{s}{Q^2}\right)^{\bar\alpha_s\chi\left(\nu\right)
+\bar\alpha_s^2\chi^{\left(1\right)}\left(\nu\right)}
\nonumber \\
&\times&\left[\frac{\left(9+4\nu^2\right)}{\nu\left(1+\nu^2\right)}\frac{ \sinh\left(\pi\nu\right)}
{\cosh^2(\pi\nu)}\right]^2
\left[1+\frac{\alpha_s}{\pi}+\frac{\bar\alpha_s}{2}{\cal F}_1\left(\nu\right)
\right] \nonumber \\
&\times&\left[\frac{11+12 \nu^2}{9+4 \nu^2}\left(1+\frac{\alpha_s}
{\pi}+\frac{\bar\alpha_s}{2}{\cal F}_2(-\nu)\right)\right.
 \nonumber \\
&-&\left. \left( 1+\frac{\alpha_s}{\pi}+\frac{\bar\alpha_s}{2}{\cal F}_1\left(-\nu\right)\right)
\right]\left\lbrace1+\bar\alpha_s\Re\left[F\left(\nu\right)\right]
\right\rbrace\;,
\end{eqnarray}

\begin{eqnarray}
\sigma^{\rm (CK)}_{LT}
&=&\left( \sum_q e_q^2\right)^2\frac{\alpha^2
\alpha_s^2}{Q^2}\frac{\pi^2}{2^8} \int_{-\infty}^{+\infty} d\nu
\left(\frac{s}{Q^2}\right)^{\bar\alpha_s\chi\left(\nu\right)
+\bar\alpha_s^2\chi^{\left(1\right)}\left(\nu\right)}
\nonumber \\
&\times&\left[\frac{\left(9+4\nu^2\right)}{\nu\left(1+\nu^2\right)}\frac{ \sinh\left(\pi\nu\right)}
{\cosh^2(\pi\nu)}\right]^2
\left[1+\frac{\alpha_s}{\pi}+\frac{\bar\alpha_s}{2}{\cal F}_1\left(-\nu\right)
\right] \nonumber \\
&\times&\left[\frac{11+12 \nu^2}{9+4 \nu^2}\left(1+\frac{\alpha_s}
{\pi}+\frac{\bar\alpha_s}{2}{\cal F}_2(\nu)\right)\right.
 \nonumber \\
&-&\left. \left( 1+\frac{\alpha_s}{\pi}+\frac{\bar\alpha_s}{2}{\cal F}_1\left(\nu\right)\right)
\right]\left\lbrace1+\bar\alpha_s\Re\left[F\left(\nu\right)\right]
\right\rbrace\;,
\end{eqnarray}
where $\alpha_s=\alpha_s(Q)$, ${\cal F}_{1,2}(\nu)$ and $\Re[F(\nu)]$ are given in Eq.~(52)
of~\cite{Balitsky2012} and in Eq.~(3.37) of~\cite{Chirilli2014}, respectively.

\begin{figure}[tb]
\centering
\includegraphics[scale=1]{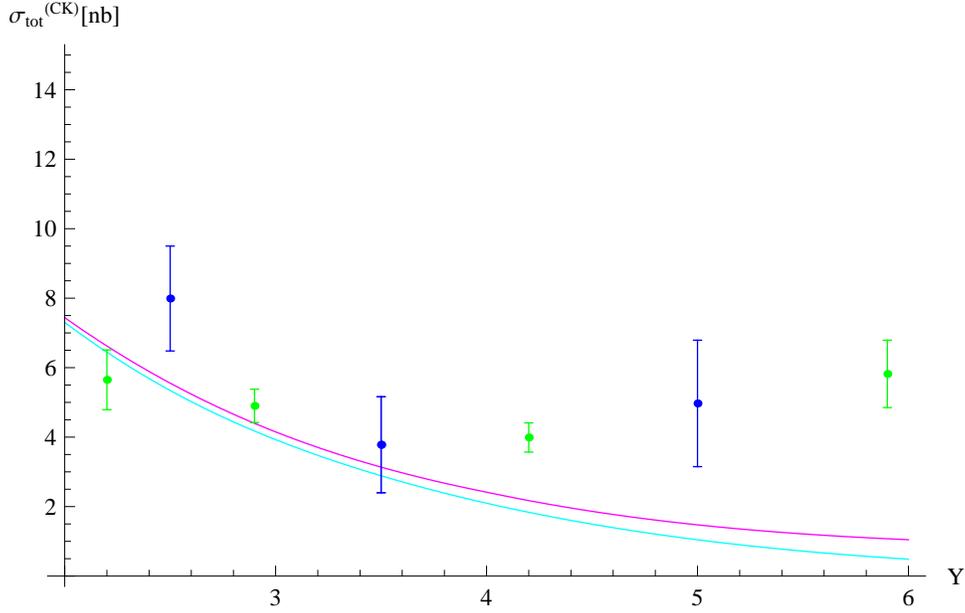}
\caption[]{$\sigma^{\rm (CK)}_{\rm tot}$ {\it versus} $Y$ at
$Q^2=17$ GeV$^2$ ($n_f=4$) (magenta line), together with the experimental
data from OPAL (blue points, $Q^2=18$ GeV$^2$) and L3 (green points,
$Q^2=16$ GeV$^2$); the cyan line represents the LO quark box
contribution only.}
\label{chirilli}
\end{figure}

In Fig.~\ref{chirilli} we report the behavior of $\sigma^{\rm (CK)}_{\rm tot}$
with $Y\equiv\ln(s/Q^2)$ for $Q^2$=17 GeV$^2$ with $n_f=4$ and contrast it
with the experimental data from CERN LEP2, namely three data points from
OPAL~\cite{Abbiendi:2001tv} ($Q^2$=18~GeV$^2$) and four data points
from L3~\cite{Achard:2001kr} ($Q^2$=16~GeV$^2$).
We see that the original Chirilli-Kovchegov representation for the cross
section (at natural values of the scales, $s_0=\mu_R^2=Q^2$) gives a very
small BFKL contribution and does not agree well with data above $Y=4$.

In the following subsections, we are going to consider other representations
of the cross section, equivalent to the Chirilli-Kovchegov one within the NLA,
and admit the possibility of moving the energy scale $s_0$ and the
renormalization scale $\mu_F$ from the ``natural'' kinematic value to some
``optimal'' scales, determined according the PMS or the BLM methods.

\subsection{Series representation with PMS optimization}

A convenient representation of the total cross section is the so-called
``series representation'', already used in Refs.~\cite{mesons_1-2,mesons_3},
which has the advantage of making manifest the BFKL resummation of leading
and subleading energy logarithms and is very practical in numerical
computations. It consists in writing the total cross section as follows
\beq
\sigma^{\rm (series)}_{\rm tot} (s,Q)
= \sigma^{\rm (series)}_{TT} + \sigma^{\rm (series)}_{LL}
+ \sigma^{\rm (series)}_{TL} + \sigma^{\rm (series)}_{LT}
+ \sigma_{\rm LO \ box}\;,
\label{sigma_series}
\eeq
where for $i,k=L,T$
\begin{eqnarray}
Q^2 \sigma^{\rm (series)}_{ik}&=&\frac{1}{\left(2\pi\right)^2}
\left\lbrace b_0^{ik}+\sum_{n=1}^\infty\bar\alpha_s^n(\mu_R) 
\ b_n^{ik} \left[\left(Y-Y_0 \right)^n+d_n^{ik}(s_0,\mu_R)
\left(Y-Y_0 \right)^{n-1}\right]\right\rbrace\;,
\end{eqnarray}
with $Y_0\equiv\ln(s_0/Q^2)$ and
\beq
b_n^{ik}=\int_{-\infty}^{+\infty}d\nu F_i\left(\nu\right)F_k
\left(-\nu\right)\frac{\chi^n\left(\nu\right)}{n!} \;,
\eeq
\begin{eqnarray}
d_n^{ik}&=&n\ln\frac{s_0}{Q^2}+\frac{\beta_0}{4 N_c}
\left[\frac{b_{n-1}^{ik}}{b_n^{ik}}\left(\left(n+1\right)
\ln\frac{\mu_R^2}{Q^2}+\frac{5}{3}\left(n-1\right)\right)
-\frac{n\left(n-1\right)}{2}\right]
\\
\nonumber
&+&\frac{1}{b_n^{ik}}\int_{-\infty}^{+\infty}d\nu \, F_i\left(\nu\right)
F_k\left(-\nu\right)\left[\frac{\chi^{n-1}\left(\nu\right)}
{\left(n-1\right)!}\left(\frac{\bar F_i^{(1)}(\nu)}
{ F_i(\nu)}+\frac{\bar F_k^{(1)}(-\nu)}{F_k(-\nu)} \right)
+\frac{\chi^{n-2}\left(\nu\right)}{\left(n-2\right)!}
\bar\chi\left(\nu\right)\right]\;,
\end{eqnarray}
where we denoted for shortness $ \bar F_i^{(1)}(\nu)\equiv F_i^{(1)}\left(\nu,s_0=Q^2,\mu_R=Q\right)$.

\begin{figure}[tb]
\centering
\includegraphics[scale=1]{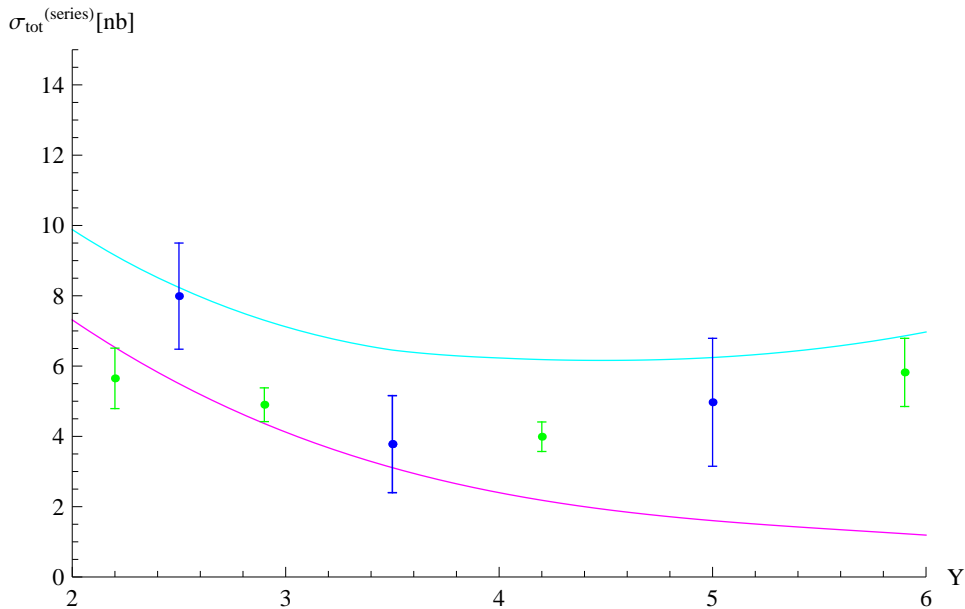}
\caption[]{$\sigma^{\rm (series)}_{\rm tot}$ {\it versus} $Y$ at
$Q^2=17$ GeV$^2$ ($n_f=4$) (magenta line), together with the experimental
data from OPAL (blue points, $Q^2=18$ GeV$^2$) and L3 (green points,
$Q^2=16$ GeV$^2$); the cyan line represents the result of
Ref.~\cite{Caporale2008} (see Fig.~3 there).}
\label{seriesPMS}
\end{figure}

Our results for $\sigma^{\rm (series)}_{\rm tot}$ at $Q^2$=17 GeV$^2$,
obtained after truncation of the series at $n=40$, are summarized in Table~\ref{tab:seriesPMS}, where
we report, for each of the $Y$ values considered, also the optimal values
of the energy scale $Y_0$ and renormalization scale $\mu_R$ found by
the PMS method. In Fig.~\ref{seriesPMS} we compare an interpolation of
the data given in Table~\ref{tab:seriesPMS} with the experimental data
from LEP2 and with the result obtained in Ref.~\cite{Caporale2008} by the
same method, but in the approximation where LO photon impact factors were
used instead of NLO ones ({\it i.e.} the same approach as here, but with
$\bar F_i^{(1)}(\nu)\to 0$). We observe that the large optimal values for the
scales we find in this approach lead to a very low contribution to the
cross section from the BFKL resummation and the overall scenario is basically
the same as for the Chirilli-Kovchegov representation. We note that
the big difference between this and the approximated result obtained in
Ref.~\cite{Caporale2008} is a clear indication that the effect of NLO
corrections to the impact factors is very substantial.

\begin{table}[tb]
\centering
\caption{Values of $\sigma^{\rm (series)}_{\rm tot}$ for several values of
$Y$ at $Q^2=17$ GeV$^2$; the last two columns give the optimal values of the
renormalization and energy scales.}
\vspace{0.4cm}
\begin{tabular}{llcc}
\hline\noalign{\smallskip}
$Y$ & $\sigma^{\rm (series)}_{\rm tot}$[nb] & $\mu_R/Q$ & $Y_0$ \\
\noalign{\smallskip}\hline\noalign{\smallskip}
2    & 7.3141 & 18 & 1 \\
3.5  & 3.1095 & 10 & 3 \\
4.5  & 1.9187 & 10 & 4 \\
6    & 1.1909 & 16 & 5 \\
\noalign{\smallskip}\hline
\end{tabular}
\label{tab:seriesPMS}
\end{table}

\subsection{Exponential representation with PMS optimization}

Here we consider representations of the NLA total cross section where the
NLO corrections to the kernel are exponentiated, in two options, which
differ by a subleading term given by the product of the two NLO corrections
of the photon impact factors:
\beq
\sigma^{\rm (exp, \ 1)}_{\rm tot} (s,Q)
= \sigma^{\rm (exp, \ 1)}_{TT} + \sigma^{\rm (exp, \ 1)}_{LL}
+ \sigma^{\rm (exp, \ 1)}_{TL} + \sigma^{\rm (exp, \ 1)}_{LT}
+ \sigma_{\rm LO \ box}\;,
\label{sigma_exp_1}
\eeq
and
\beq
\sigma^{\rm (exp, \ 2)}_{\rm tot} (s,Q)
= \sigma^{\rm (exp, \ 2)}_{TT} + \sigma^{\rm (exp, \ 2)}_{LL}
+ \sigma^{\rm (exp, \ 2)}_{TL} + \sigma^{\rm (exp, \ 2)}_{LT}
+ \sigma_{\rm LO \ box}\;,
\label{sigma_exp_2}
\eeq
with
\begin{eqnarray}
\label{sigma_exp_11}
\sigma^{\rm (exp, \ 1)}_{ik}
&=&\frac{1}{\left(2\pi\right)^2 Q^2}
\int_{-\infty}^{+\infty}d\nu\ e^{\left(Y-Y_0\right)
\left[\bar\alpha_s\left(\mu_R\right)\left(1+\frac{\bar\alpha_s\left(\mu_R\right)\beta_0}{4N_c}\ln\frac{\mu_R^2}{Q^2}\right)\chi\left(\nu\right)
+\bar\alpha_s^2\left(\mu_R\right)\chi^{\left(1\right)}
\left(\nu\right)\right]} \nonumber \\
&\times& F_i(\nu) F_k(-\nu) \left[1+\bar\alpha_s\left(\mu_R\right)
\left(\frac{F_i^{(1)}(\nu)}{F_i(\nu)}
+\frac{F_k^{(1)}(-\nu)}{F_k(-\nu)}\right)\right]
\end{eqnarray}
and
\begin{eqnarray}
\label{sigma_exp_22}
\sigma^{\rm (exp, \ 2)}_{ik}
&=&\frac{1}{\left(2\pi\right)^2 Q^2}
\int_{-\infty}^{+\infty}d\nu\ e^{\left(Y-Y_0\right)
\left[\bar\alpha_s\left(\mu_R\right)\left(1+\frac{\bar\alpha_s\left(\mu_R\right)\beta_0}{4N_c}\ln\frac{\mu_R^2}{Q^2}\right)\chi\left(\nu\right)
+\bar\alpha_s^2\left(\mu_R\right)\chi^{\left(1\right)}
\left(\nu\right)\right]}  \nonumber \\
&\times& F_i(\nu) F_k(-\nu) \left[1+\bar\alpha_s\left(\mu_R\right)
\left(\frac{F_i^{(1)}(\nu)}{F_i(\nu)}+\frac{F_k^{(1)}(-\nu)}{F_k(-\nu)}\right)
\right. \nonumber \\
&+& \left. \bar\alpha_s^2\left(\mu_R\right)\left(\frac{F_i^{(1)}(\nu)}
{F_i(\nu)} \frac{F_k^{(1)}(-\nu)}{F_k(-\nu)}\right)\right]\, .
\end{eqnarray}
In these equations we denote for shortness $F_{i,k}^{(1)}(\nu)\equiv F_{i,k}^{(1)}(\nu, s_0,\mu_R)$.

\begin{figure}[tb]
\centering
\includegraphics[scale=1]{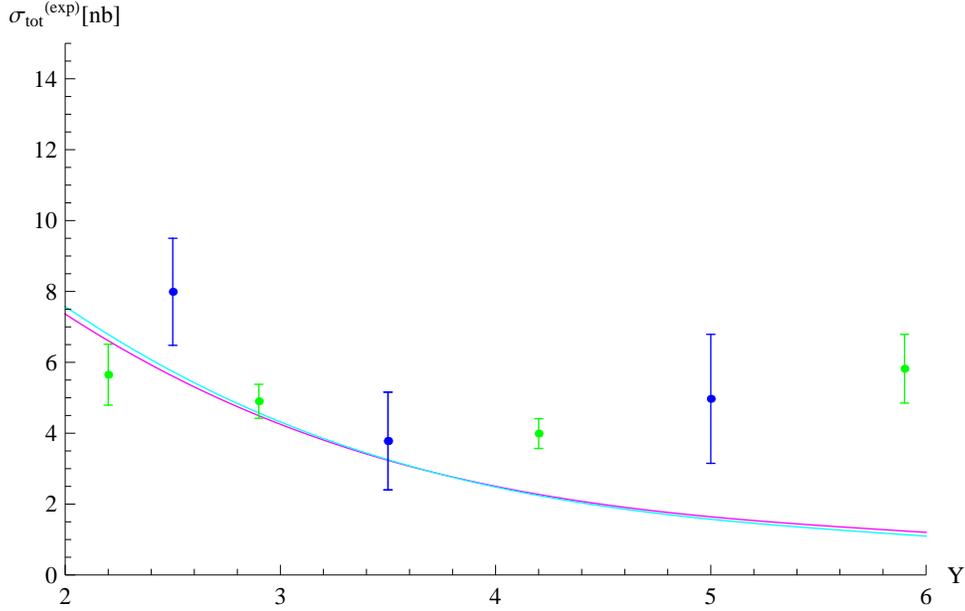}
\caption[]{$\sigma^{\rm (exp, \ 1)}_{\rm tot}$ (magenta line)
and $\sigma^{\rm (exp, \ 2)}_{\rm tot}$ (cyan line) {\it versus} $Y$
at $Q^2=17$ GeV$^2$ ($n_f=4$), together with the experimental
data from OPAL (blue points, $Q^2=18$ GeV$^2$) and L3 (green points,
$Q^2=16$ GeV$^2$).}
\label{expPMS}
\end{figure}

We used these two exponentiated representations together with the PMS
method to fix the values of the energy scales and obtained the results
given in Table~\ref{tab:expPMS}. We can see that the variant (2) of the
exponentiated cross section gets lower values for the optimal energy scales,
thus implying that the inclusion of the subleading term with the product
of the NLO impact factors catches a relevant part of the unknown next-to-NLA
corrections. However, as shown in Fig.~\ref{expPMS}, the absolute value
of the cross section remains low and undershoots LEP2 data substantially
in the same fashion as the two previous representations.

\begin{table}
\centering
\caption{Values of $\sigma^{\rm (exp, \ 1,2)}_{\rm tot}$ for several values of
$Y$ at $Q^2=17$ GeV$^2$; the columns 3-4 and 6-7 give the
optimal values of the renormalization and energy scales.}
\vspace{0.4cm}
\begin{tabular}{llcclcc}
\hline\noalign{\smallskip}
$Y$ & $\sigma^{\rm (exp, \ 1)}_{\rm tot}$[nb] & $\mu_R/Q$ & $Y_0$
    & $\sigma^{\rm (exp, \ 2)}_{\rm tot}$[nb] & $\mu_R/Q$ & $Y_0$ \\
\noalign{\smallskip}\hline\noalign{\smallskip}
2   & 7.36281 & 18 & 1 & 7.57706 & 8 & 1 \\
3.5 & 3.23512 & 18 & 3 & 3.25243 & 8 & 1 \\
4.5 & 1.98923 & 18 & 4 & 1.9419  & 8 & 1 \\
6   & 1.20222 & 18 & 5 & 1.09588 & 8 & 1 \\
\noalign{\smallskip}\hline
\end{tabular}
\label{tab:expPMS}
\end{table}

\subsection{Exponential representation with BLM optimization}

Here we consider the first variant of the exponentiated cross section discussed
in the previous subsection, combined with two different implementations
(variants $(a)$ and $(b)$) of the BLM method (for a justification of the
formulas below, we refer to~\cite{us_BLM}):
\beq
\sigma^{\rm (BLM, \ a)}_{\rm tot} (s,Q)
= \sigma^{\rm (BLM, \ a)}_{TT} + \sigma^{\rm (BLM, \ a)}_{LL}
+ \sigma^{\rm (BLM, \ a)}_{TL} + \sigma^{\rm (BLM, \ a)}_{LT}
+ \sigma_{\rm LO \ box}\;,
\label{sigmaBLM_a}
\eeq
where
\begin{eqnarray}
\label{BLMa}
\sigma_{ik}^{\rm (BLM, \ a)}
&=&\frac{1}{\left(2\pi\right)^2 Q^2}\int_{-\infty}^{+\infty} d\nu\ e^{\left(Y-Y_0\right)
\left[ {\bar{\alpha}}_s\left(\mu_{R, a}^{\rm BLM}\right)\chi\left(\nu\right)
+ \left({\bar{\alpha}}_s\left(\mu_{R, a}^{\rm BLM}\right)\right)^2
\left(-\frac{T^{\beta}}{N_c}\chi\left(\nu\right)
+{\bar{\chi}}\left(\nu\right)-\frac{\beta_0}{8N_c}\chi^2\left(\nu\right)
\right)\right]} \nonumber \\
&\times&F_i(\nu) F_k(-\nu)
\left[ 1
+{\bar{\alpha}}_s\left(\mu_{R, a}^{\rm BLM}\right)
\left(\frac{\tilde F_i^{(1)}(\nu)}{F_i(\nu)}+ \frac{\tilde F_k^{(1)}(-\nu)}{F_k(-\nu)}-2\frac{T^{\beta}}{N_c}\right)
\right]\, ,
\end{eqnarray}
with
\beq
\left(\mu_{R, a}^{\rm BLM}\right)^2=Q^2\ \exp\left[2\left(1+\frac{2}{3}I\right)
-\frac{5}{3}\right]\;,
\eeq
and
\beq
\sigma^{\rm (BLM, \ b)}_{\rm tot} (s,Q)
= \sigma^{\rm (BLM, \ b)}_{TT} + \sigma^{\rm (BLM, \ b)}_{LL}
+ \sigma^{\rm (BLM, \ b)}_{TL} + \sigma^{\rm (BLM, \ b)}_{LT}
+ \sigma_{\rm LO \ box}\;,
\label{sigmaBLM_b}
\eeq
where
\begin{eqnarray}
\label{BLMb}
\sigma_{ik}^{\rm (BLM, \ b)}
&=&\frac{1}{\left(2\pi\right)^2Q^2}\int_{-\infty}^{+\infty} d\nu
\ e^{\left(Y-Y_0\right)\left[ {\bar{\alpha}}_s\left(\mu_{R, b}^{\rm BLM}\right)
\chi\left(\nu\right)+\left({\bar{\alpha}}_s\left(\mu_{R, b}^{\rm BLM}\right)
\right)^2
\left(-\frac{T^{\beta}}{N_c}\chi\left(\nu\right)+{\bar{\chi}}\left(\nu\right)
\right) \right]} \nonumber \\
\nonumber
&\times&  F_i(\nu) F_k(-\nu)\left[ 1+{\bar{\alpha}}_s\left(\mu_{R, b}^{\rm BLM}\right)
\left(\frac{\tilde F_i^{(1)}(\nu)}{F_i(\nu)} + \frac{\tilde F_k^{(1)}(-\nu)}{F_k(-\nu)}
\right)\right. \\
&+&{\bar{\alpha}}_s\left(\mu_{R, b}^{\rm BLM}\right)
\left( \left.-\frac{2T^{\beta}}{N_c}+\frac{\beta_0}{4N_c}\chi\left(\nu\right)
\right)\right]\;,
\end{eqnarray}
with
\beq
\left(\mu_{R, b}^{\rm BLM}\right)^2=Q^2\ \exp\left[2\left(1+\frac{2}{3}I\right)
-\frac{5}{3}+\frac{1}{2}\chi\left(\nu\right)\right]\;.
\eeq
In Eqs. (\ref{BLMa}) and (\ref{BLMb}) the LO impact factors have to be evaluated with $\alpha_s=\alpha_s\left(\mu_{R, a}^{\rm BLM}\right)$ and $\alpha_s=\alpha_s\left(\mu_{R, b}^{\rm BLM}\right)$ respectively. In both cases, we have
\[
T^\beta= -\frac{\beta_0}{2} \left[1 + \frac{2}{3}I \right] \;, \;\;\;\; I
\simeq 2.3439 \, .
\]
Besides, in Eqs. (\ref{BLMa}) and (\ref{BLMb}) we denote
\beq
\frac{\tilde F_i^{(1)}(\nu)}{F_i(\nu)}\equiv \frac{ F_i^{(1)}(\nu,s_0,\mu_R)}{F_i(\nu)}-\frac{\beta_0}{4N_c}\left(\ln\frac{\mu_R^2}{Q^2} +\frac{5}{3}\right)\, .
\eeq

\begin{figure}[tb]
\centering
\includegraphics[scale=1]{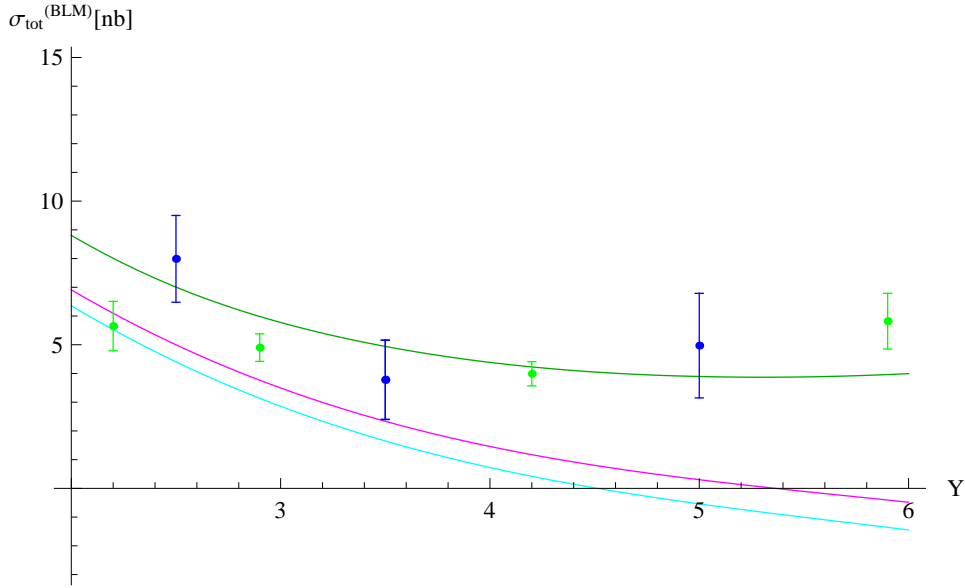}
\caption[]{$\sigma^{\rm (BLM, \ a)}_{\rm tot}$ (cyan line)
and $\sigma^{\rm (BLM, \ b)}_{\rm tot}$ (magenta line) {\it versus} $Y$
at $Q^2=17$ GeV$^2$ ($n_f=4$), together with the experimental
data from OPAL (blue points, $Q^2=18$ GeV$^2$) and L3 (green points,
$Q^2=16$ GeV$^2$); the green line represents the result of
Ref.~\cite{Caporale2008} (see Fig.~3 there).}
\label{expBLM}
\end{figure}

The results in this approach, calculated at $s_0=Q^2$, are shown in
Fig.~\ref{expBLM}, where we can see that the cross section is very low and
starts even to be negative at larger values of $Y$. In the same Figure,
we show also the result obtained in
Ref.~\cite{Caporale2008} by a similar approach, but in the approximation
where photon impact factors were taken at the LO.

\section{Discussion}

In this paper we have studied the $\gamma^* \gamma^*$ total cross section
in the NLA BFKL approach. First we have extracted the NLO corrections to
the photon impact factor from two recent
papers~\cite{Balitsky2012,Chirilli2014}, then we have used them to build
several representations of the total cross section, equivalent within the NLA,
but taking into account in a different way pieces of the (unknown)
subleading contributions. We have combined these different representations
with two among the most common methods for the optimization of a
perturbative series, namely PMS and BLM, and compared their behavior with
the energy with the only available experimental data, those from the LEP2
collider. We have considered also the numerical implementation of formulas
describing the BFKL contribution to $\gamma^* \gamma^*$ total cross section,
derived originally by Chirilli and Kovchegov~\cite{Chirilli2014}.

We have found that, in general, the effect of the BFKL resummation is small
and changes only by little the determination coming from the LO quark box
diagrams. This means that, in the considered range of energies, the NLO
corrections to the photon impact factor compensate almost exactly the LO
ones. Indeed, previous estimates of the cross
section~\cite{Brodsky:2002ka,Brodsky:1998kn,Caporale2008, Zheng} using LO
impact factors together with the NLA BFKL Green's function showed a better
agreement with LEP2 data.

In other words, the account of the Balitsky and Chirilli expression for
NLO photon impact factor reduces the BFKL contribution to the cross section
to very small values making it impossible to describe LEP2 data as a sum of
BFKL and LO QED quark box contributions. Note that, the LO QED quark box
itself receives, at higher QCD orders, large corrections enhanced by double
logs. Their resummation is important and leads to a considerable enhancement
of the quark box contribution -- see Ref.~\cite{BL03} for details,
but still these effects are not large enough for a good description of LEP2
data at $Y=3.5\div 6$ without a sizable BFKL contribution.

There could be many reasons for this problem at $Y=3.5\div 6$. The first,
obvious one, is that even at such high energies the BFKL contribution could
be still not the dominant one in comparison with terms which
are suppressed by powers of the energy $\sim 1/s$, and are not included in
the present consideration. In particular, terms, subleading in energy,
coming from diagrams with gluon exchange in the $t$-channel, see
Fig.~\ref{fig:BFKL}, can be important.
We could also argue that the presumably large effects in the next-to-NLA are
not reduced under enough satisfactory control by the representations of the
cross section and by the optimization methods we have considered in this work.
In this respect, it would be interesting to test also approaches based on
collinear improvement~\cite{collinear}. However, the consideration of these
issues goes beyond the scope of present paper.

\begin{figure}[tb]
\centering
\includegraphics[scale=0.64]{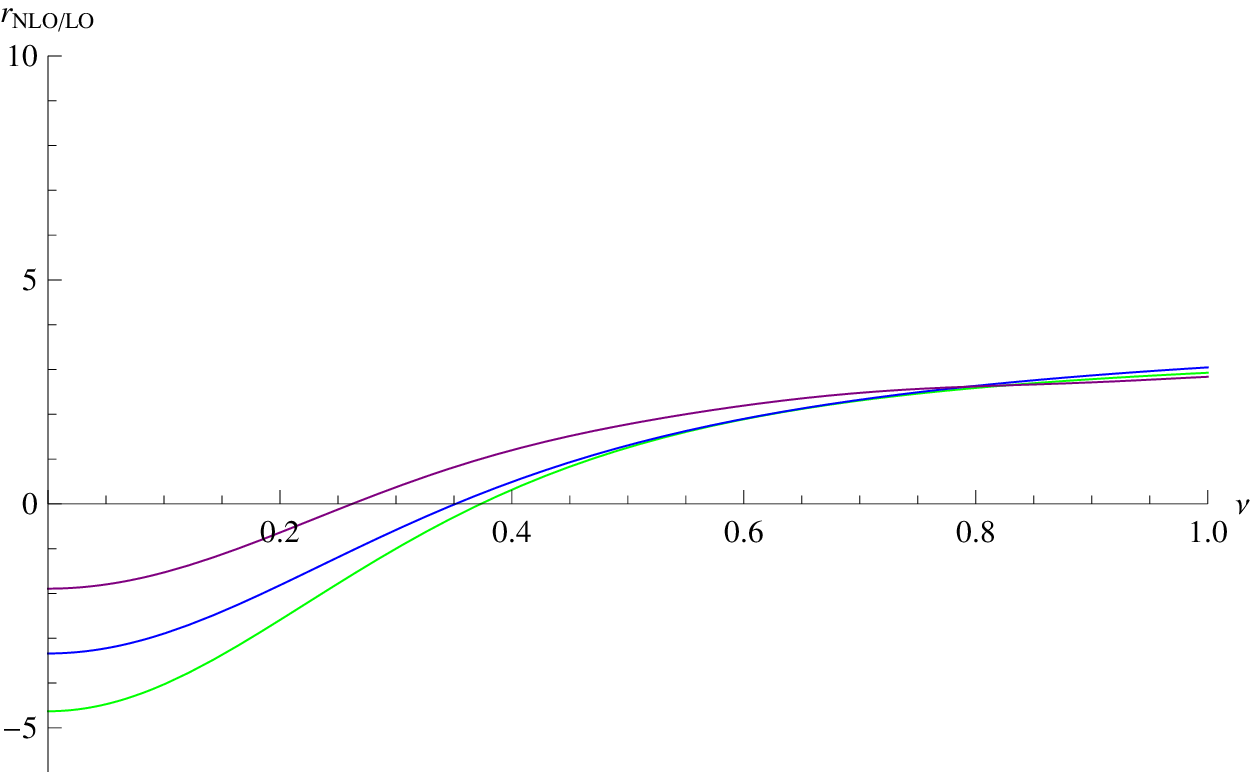}
\includegraphics[scale=0.64]{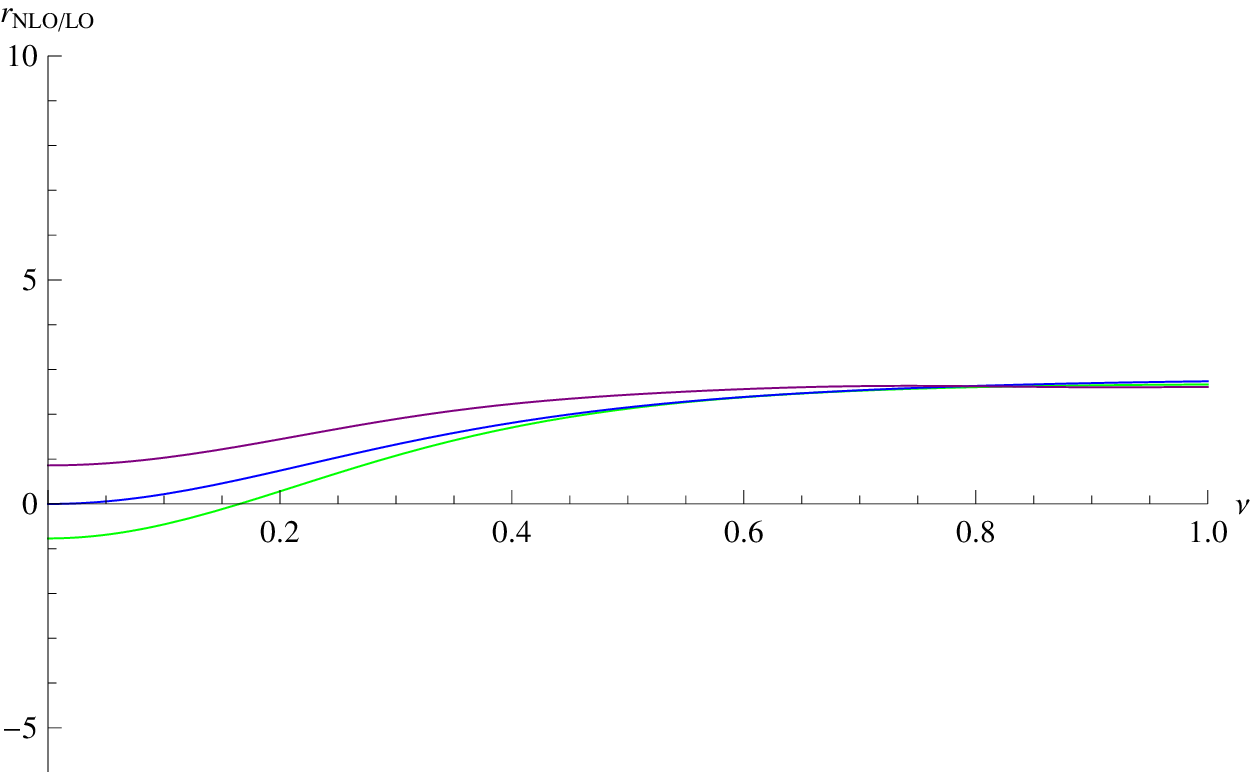}
\caption[]{Behavior of $r_{\rm NLO/LO}^T(\nu,s_0,\mu_R)$ (green),
$r_{\rm NLO/LO}^L(\nu,s_0,\mu_R)$ (blue) and $r_{\rm NLO/LO}^{(\rm mesons)}(\nu,s_0,\mu_R)$
(violet) for the following cases:
$Q^2=\mu_R^2=17$ GeV$^2$, $Y_0=0$ on the left and
$Q^2=17$ GeV$^2$, $\mu_R^2=(10Q)^2$, $Y_0=2.2$ on the right.}
\label{photons_vs_mesons}
\end{figure}

Definitely, the problems with our description of LEP2 data in the present
context originate from the large negative value of NLO contributions to the
photon impact factor. For this reason, we will discuss several issues
related with this quantity. First, we want to illustrate our statement that
the NLO corrections to the photon impact factor turned to be very large.
For this purpose we plot in Fig.~\ref{photons_vs_mesons} the factors which,
in the case of transverse and longitudinal photon polarizations, control
the normalization of the cross section in the case of the exponential
representation~(\ref{sigma_exp_11}),
\beq
r_{\rm NLO/LO}^{(T,L)}(\nu,s_0,\mu_R)\equiv
1+\bar\alpha_s(\mu_R) \left(\frac{F_{T,L}^{(1)}(\nu,s_0,\mu_R)}{F_{T,L}(\nu)}
+\frac{F_{T,L}^{(1)}(-\nu,s_0,\mu_R)}{F_{T,L}(-\nu)}\right) \;.
\eeq
For the sake of comparison, we present in Fig.~\ref{photons_vs_mesons}
also the similar quantity $r_{\rm NLO/LO}^{(\rm mesons)}$ which appeared in
the description of the process $\gamma^* \gamma^*$ to two light vector mesons,
see Refs.~\cite{mesons_1-2,mesons_3},
\beq
r_{\rm NLO/LO}^{(\rm mesons)}(\nu,s_0,\mu_R)\equiv 1+
\bar\alpha_s(\mu_R) \left(\frac{c_1^{(1)}(\nu,s_0,\mu_R)}{c_1(\nu)}
+\frac{c_2^{(1)}(-\nu,s_0,\mu_R)}{c_2(-\nu)}\right)\; .
\eeq
The $\nu$ dependence of these quantities is shown on the left panel of
Fig.~\ref{photons_vs_mesons} in the case of natural scale choice,
$s_0=\mu_R^2=Q^2$, whereas on the right panel we show the same quantities
calculated for $\mu_R^2=(10Q)^2$, $Y_0=2.2$, the values of scales which were
obtained in Refs.~\cite{mesons_1-2,mesons_3} during PMS optimization
procedure applied to $\gamma^*\gamma^*\to VV$ process.
In the region of large-$\nu$ the results are similar in all the three
cases, whereas in the low-$\nu$ region they differ substantially.
For natural scales (left panel) and $\nu\le 0.25$ all the three quantities are
negative; note that it is the region of $\nu$ that dominates the
$\nu$-integral appearing in the cross section. We see that in this
$\nu$-region the NLO corrections to the impact factors are negative and turn
to be much larger for $\gamma^*\to\gamma^*$ (especially in the case of
transverse polarization) in comparison to the case of $\gamma^*\to V$ impact
factor. Such a difference remains to be understood.

The impact factors in BFKL approach depend on the scales $s_0$ and $\mu_R$,
see Eqs.~(\ref{ifT}) and~(\ref{ifL}). Comparing the left and right panels of
Fig.~\ref{photons_vs_mesons}, one can see that this effect is important.
In particular, $\gamma^*\to V$ and $\gamma_L^*\to\gamma^*_L$ impact factors
become positive in the whole $\nu$ range when one goes from the natural choice
of scales to $\mu_R^2=(10Q)^2$, $Y_0=2.2$. But it is not the case for
the $\gamma_T^*\to\gamma^*_T$ impact factor, which remains negative-valued
for a substantial range of small $\nu$. Note that the transverse polarization
gives the most important contributions ($\sigma_{TT}$) to the effective
$\gamma^*\gamma^*$ cross section which we consider in this paper. This
observation explains, on the qualitative level, the very high values of
optimal scales in Tables~\ref{tab:seriesPMS} and~\ref{tab:expPMS},
which we found with PMS method for the $\gamma^*\gamma^*$ total cross section.

The other issue we want to mention here is the color structure of the NLO
parts of the photon impact factors.
We observe that the NLO impact factors as extracted from~\cite{Chirilli2014}
have very simple subleading $\sim 1/N_c^2$ contributions, which appear only
in the trivial third terms of Eqs.~(\ref{ifT}) and~(\ref{ifL}). This is in
sharp contrast with what happens in the case of the NLO virtual photon to
light vector meson impact factor~\cite{IKP04} and of the NLO forward jet
impact factor~\cite{MN_IF}.
It would be interesting to understand the reason for the practically
complete cancellation of the subleading $1/N_c^2$ terms which takes place here.

\begin{figure}[tb]
\centering
\includegraphics[scale=0.64]{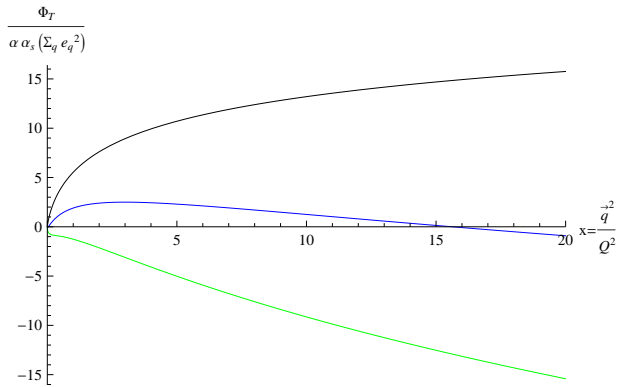}
\caption[]{Behavior of the photon impact factor (the transverse polarization) with the Reggeon transverse
momentum $\vec q$, through the variable $x\equiv \vec q^{\, 2}/Q^2$.
The black curve represents the LO impact factor, the green curve the sum
of LO and NLO parts derived from Eq.~(\ref{ifT}) and the blue curve the
same as the green curve, but with the NLO part reduced by the factor 1.87.}
\label{Bartels-Chachamis}
\end{figure}

Finally, we want to compare the results for the photon impact factor which we
used in this paper (derived from the results
in~\cite{Balitsky2012,Chirilli2014})
with the ones obtained in the conventional BFKL approach by Bartels and
collaborators~\cite{gammaIF}. Unfortunately, some information (in numerical
form) about the final result for the impact factor is available
only for the case of transverse polarization -- see
Ref.~\cite{Chachamis:2006zz} in the ``Diffraction 2006'' workshop proceedings.
To make such a comparison we need to transfer the photon impact factor from
the $\nu$- to the transverse momentum representation:
\beq
\Phi_T(x,s_0,\mu_R)= \int\limits^{\infty}_{-\infty}d\nu \frac{(x)^{-i \nu + {1\over 2}}}{\pi \sqrt{2}}\left[F_{T}(\nu)+\bar\alpha_s(\mu_R) F_{T}^{(1)}(\nu,s_0,\mu_R)\right]\, ,
\label{x-rep}
\eeq
where the variable $x$ is defined as a dimensionless ratio of the Reggeon
transverse momentum $\vec q$ and the photon virtuality squared: $x\equiv
\vec q^{\, 2}/Q^2$.
The plot of $\Phi_T(x,s_0,\mu_R)$ as a function of $x$ is presented in
Fig.~6 of Ref.~\cite{Chachamis:2006zz}. Here we perform the $\nu$ integration
in~(\ref{x-rep}) using Eq.~(\ref{ifT}), the Balitsky-Chirilli result for the
transverse photon impact factor transformed to the conventional BFKL
scheme.  In our Fig.~\ref{Bartels-Chachamis} we
present results for $\Phi_T(x,s_0,\mu_R)/(\alpha \, \alpha_s
\left( \sum_q e_q^2 \right))$, where we used the following settings in order
to compare with~\cite{Chachamis:2006zz}: $s_0=10$ GeV$^2$ , $Q^2=\mu_R^2=15$
GeV$^2$; moreover, we take $n_f=1$ and $\alpha_s=0.177206$.\footnote{We are
grateful to Grigorios Chachamis who provided us with the information about
the numeric values of $n_f$ and $\alpha_s$ which were used to produce
Fig.~6 of~\cite{Chachamis:2006zz}.} In Fig.~\ref{Bartels-Chachamis} we show
the behavior of the photon impact factor with the Reggeon transverse momentum
$\vec q$, through the variable $x$. In Fig.~\ref{Bartels-Chachamis} the black
curve represents the LO impact factor,  the green curve gives LO plus NLO
parts derived from Eq.~(\ref{ifT}), and the blue curve LO plus NLO parts derived from Eq.~(\ref{ifT}),
when NLO contribution is reduced by the factor 1.87.  We see that the NLO corrections
are rather large and it is clear that the $x$-shape of $\Phi(x,s_0,\mu_R)$ is
rather sensitive to their value.
Comparing the shape of the $x$-dependence in Fig.~6 of~\cite{Chachamis:2006zz}
with the NLO curves in Fig.~\ref{Bartels-Chachamis}, we should conclude that
the results of Balitsky and Chirilli are not in agreement with those
presented in~\cite{Chachamis:2006zz}. Interestingly, a qualitative agreement
for the $x$-shape of $\Phi(x,s_0,\mu_R)$ could be obtained only
reducing the NLO result given in Eq. ~(\ref{ifT}) by the factor $\sim 1.87$.

To summarize this discussion we would like to stress that it would be very
important if the authors of~\cite{Chachamis:2006zz} could finally publish
their results for the photon impact factor, since it would be an independent
test of the results obtained by Balitsky and Chirilli using a completely
different approach.

\vspace{1.0cm} \noindent
{\Large \bf Acknowledgments} \vspace{0.5cm}

The authors are grateful to G.~Chirilli and Yu.~Kovchegov for valuable
discussions.
D.I. thanks the Dipartimento di Fisica dell'Universit\`a della Calabria and
the Istituto Nazionale di Fisica Nucleare (INFN), Gruppo collegato di Cosenza,
for warm hospitality and financial support. The work of D.I. was also
supported in part by the grant RFBR-13-02-00695-a.
\\
The work of B.M. was supported by the European Commission, European Social
Fund and Calabria Region, that disclaim any liability for the use that can be
done of the information provided in this paper.
\\
B.M. thanks the Sobolev Institute of Mathematics
of Novosibirsk for warm hospitality during the preparation of this work.

\end{document}